# Design of narrowband infrared emitters by hybridizing guided-mode resonance structures with van der Waals materials

Mehrdad Shokooh-Saremi,[1,*] Maxime Giteau,[2] Mitradeep Sarkar,[2] and Georgia T. Papadakis[2]

[1] *Department of Electrical Engineering, Ferdowsi University of Mashhad, Mashhad 91779-48944, Iran*

[2] *ICFO - Institut de Ciències Fotòniques, The Barcelona Institute of Science and Technology, Castelldefels (Barcelona) 08860, Spain*

*Corresponding author: m_saremi@um.ac.ir*

**Abstract:** In this paper, narrowband emitters have been designed using particle swarm optimization (PSO) in the 10-20 μm infrared range. The device structure consists of an anisotropic α-MoO$_3$ layer combined with the one- and two-dimensional guided-mode resonance structures. Well-defined absorption lines are present in the reflection spectrum for both TE and TM polarizations, thereby yielding narrowband emissivity at desired wavelengths. The band structure of the designed emitters under TM polarization demonstrates distinct features unlike its TE counterpart. These features are attributed to the interaction between guided-mode resonances and phonon polaritons. The results are relevant for applications in active and passive photonic elements in mid- and long-wave IR bands.

## Introduction

Applications in thermophotovoltaic systems [1, 2], radiative cooling [3], compact optoelectronic devices [4], molecular sensing [5, 6], and infrared spectroscopy [7] require spectral control of thermal emission at mid- and long-infrared (IR) bands. In particular for applications in spectroscopy, molecular sensing and thermophotovoltaic systems [8], a narrow thermal emission spectrum is key for high efficiency and optimal device operation [9, 10].

We design narrowband mid-IR thermal emitters by coupling two distinct optical phenomena, namely guided-mode resonance (GMR) and hyperbolic phonon resonances of emerging van der Waals (vdW) polar dielectrics. The GMR effect is employed in numerous optical applications, especially for realizing functional dielectric optical elements [11-13]. GMR occurs when the incident light excites the modes of a waveguide with a periodic grating upon satisfying the grating coupling condition [12], yielding very sharp peaks in the reflection spectrum. Since GMR is a polarization-dependent mechanism, the designed emitters demonstrate polarization sensitivity, which is often a requirement. On the other hand, several vdW polar dielectrics, such as hexagonal boron nitride (hBN) and α-$MoO_3$ [14-16] possess phonons with hyperbolic dispersion, which results in various intriguing optical phenomena in the mid-IR and terahertz (THz) bands [17]. Namely, the hyperbolic phonon dispersion of these vdW crystals often leads to colossal optical anisotropy near resonance [18, 19]. The hyperbolicity of the vdW crystals provides access to electromagnetic modes with ultra-high momentum as compared to bulk isotropic materials. Such naturally hyperbolic crystals have also be employed directly or in combination with metamaterials [20, 21]. As we demonstrate below, combining inherently anisotropic van der Waals materials with GMR periodic elements results in ultra-narrow thermal emission resonances, and polarization-dependent responses.

Recent works have reported optical devices with diverse functionalities, by combining vdW polar dielectrics with periodic meta-structures. For example, Ye *et al.* proposed a structure consisting of slabs and nano-disks of α-$MoO_3$ and showed that by changing the incident polarization, in-plane phonon polaritons can be selectively excited [22]. Zheng *et al.* numerically studied the coupling between a GMR and surface phonon polaritons (SPhP) in a structure consisting of an anisotropic substrate (SiC), a dielectric spacer and a dielectric grating [23]. Pechprasarn *et al.* replaced the prism in an Otto configuration by a dielectric grating to directly excite the SPhP of the SiC substrate. They showed that a hybridization between the SPhP, GMR and Fabry-Perot (FP) modes can occur, provided that the structural parameters are chosen properly [24]. Ito *et al.* studied a coupling scheme between GMR and SPhP in a SiC substrate and concluded that this can be employed for emissivity modulation in the infrared range [25].

The aforementioned results demonstrate that introducing anisotropic materials in GMR-based optical elements may enable novel functionalities. In this paper, we leverage the high in-plane anisotropy of vdW materials with GMR to design efficient thermal emitters. We note that, until now, modeling GMRs in the presence of anisotropic materials has remained cumbersome, due to the increased number of design parameters and its inherent complexity. To address this issue, we employ an effective optimization method namely the particle swarm optimization (PSO) to design the proposed hybrid device [26, 27]. Furthermore, we explore the underlying physics of these hybrid structures to achieve narrowband emission for both linear polarization states. In a first step, we consider 1D grating to achieve narrowband emission for either TE or TM polarization. In a second step, using 2D patterning, we demonstrate narrowband emission at distinct wavelengths simultaneously for TE and TM polarizations.

**Device structure, materials and optimization method**

Figure 1(a) shows a schematic of the proposed structure. The substrate and superstrate media are KBr and air with refractive indices $n_{KBr} = 1.52$ and $n_{air} = 1.0$ in the wavelength range of interest, respectively. We consider illumination from the substrate side. Starting from the substrate, the structure comprises three layers stacked in the $z$ direction: a ZnSe layer ($n_{ZnSe} = 2.36$), a ZnSe/air grating, periodic in the $x$ direction, and a α-MoO$_3$ layer. The period of the grating and its filling factor are referred to as $\Lambda$ and F, respectively. The thicknesses of the grating, the ZnSe layer and the α-MoO$_3$ layer are d$_g$, d$_L$, and d$_M$, respectively. The structure is considered infinite in the $y$ direction. Since best quality α-MoO$_3$ are mechanically exfoliated [28], the proposed configuration is designed such that the exfoliated flakes can be directly transferred on the ZnSe grating. In addition, it has been shown that these high quality α-MoO$_3$ exfoliated flakes can retain their anisotropy for thicknesses up to a few micrometers [28].

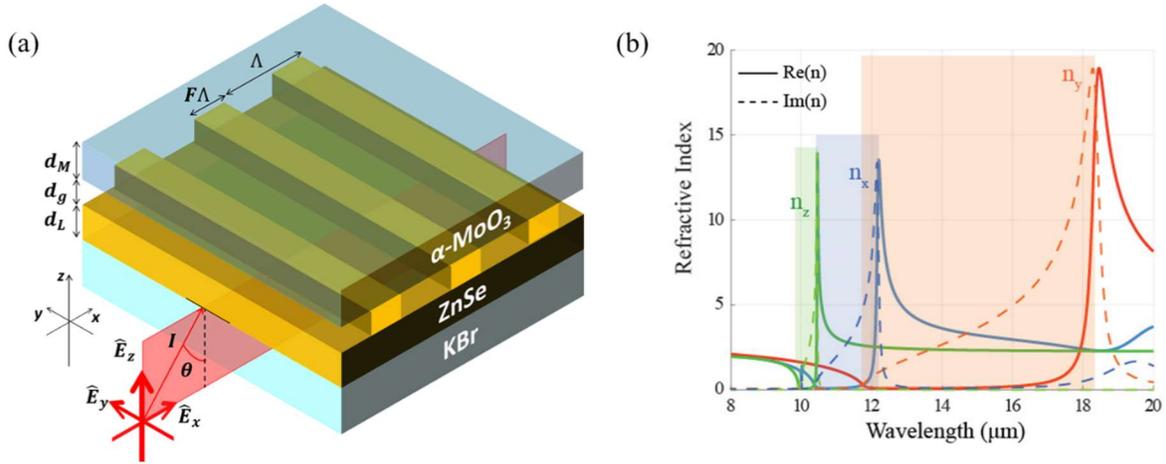

Fig. 1. (a) Schematic view of the proposed element. The period of grating, grating filling factor, thicknesses of grating, ZnSe layer and α-MoO$_3$ layer are shown by $\Lambda$, F, d$_g$, d$_L$, and d$_M$, respectively, and the structure in infinite along y axis. Illumination is from the substrate side. (b) Real and imaginary parts of refractive index ($n_i = \sqrt{\varepsilon_i}$) of α-MoO$_3$ versus wavelength, along its three principal directions. The colored areas show the extent of the corresponding Reststrahlen bands.

The orthorhombic crystal structure of α-MoO$_3$ makes it highly anisotropic. Consequently, its three lattice constants and three principal permittivities ($\varepsilon_x(\omega)$, $\varepsilon_y(\omega)$, $\varepsilon_z(\omega)$) are different, exhibiting sharp phonon modes at different frequencies in the IR range (Fig. 1(b)) [28]. The permittivities are described by a Lorentz model for coupled oscillators (three oscillators for $\varepsilon_x(\omega)$ and one oscillator for each $\varepsilon_y(\omega)$ and $\varepsilon_z(\omega)$) [28]. The corresponding Reststrahlen bands (RB), where the real part of the frequency-dependent permittivity obtains negative values, is shown in Fig. 1(b) for the three lattice directions. If the α-MoO$_3$ layer is thick enough in comparison to the skin depth defined by $\delta_i = \lambda_0/2\pi Im(\sqrt{\varepsilon_i(\lambda_0)})$, the RBs in $x$ and $y$ directions yield relatively broad-reflection bands of ~2 µm and ~6.6 µm within the wavelength range of 10-12 µm and 12-18 µm, respectively. This behavior is unique to hyperbolic media, as compared to metals that do not exhibit polarization-dependent reflection in the mid-IR. In addition, the dielectric-metal interfaces support surface plasmons, while interface of polar dielectric-dielectric supports phonon polaritons upon illumination by proper polarization.

To optimize the device with PSO, each particle of a swarm in PSO is considered as a point in an *N*-dimensional search space, which adjusts its position according to its own experience as well as the experience of other particles. For particle evaluation, a fitness function (FF) is defined and calculated for each particle according to its N optimization parameters. The position of the particles is updated iteratively until either an optimum solution is obtained or the maximum number of iterations is reached. The standard, real-coded PSO is employed in this paper [27]. Rigorous coupled-wave analysis method (RCWA) has been employed as the computational kernel to calculate the reflectance $R$ and transmittance $T$ spectra and the fitness functions [29-31]. The spectral emissivity is derived from absorptivity following Kirchhoff's law of thermal radiation, which states that for an arbitrary body emitting and absorbing thermal radiation in a thermodynamic equilibrium, the emissivity ($\mathcal{E}$) is equal to the absorptivity ($A$): $\mathcal{E} = A = 1 - R - T$. PSO is able to find the structural optimal parameters for the desired spectra without prior information or initial values. PSO is an efficient approach in comparison to semi-analytical methods, as the latter usually provide parameters that need refinements in order to meet the desired spectra, or they are used as an initial step for providing initial values to optimization methods. In addition, utilization of these methods is often limited to the simpler structure and non-dispersive materials. PSO does not require such initial values derived from semi-analytical approaches.

**Design of 1D narrowband emitters**

We consider the device configuration shown in Fig. 1(a) for normal incidence. The TE (TM) polarization refers to the incident electric (magnetic) field being perpendicular to the plane of incidence (under normal incidence, this corresponds to the electric field being along y and x axes, respectively). As the first illustration, we design a narrowband emitter for the wavelength of 15 μm (almost at the center of $RB_y$). For the PSO optimization, we define ideal performance for TE polarization such that the device exhibits unitary reflectance for wavelengths $\lambda = \{14.0, 14.5, 15.5, 16.0\}$ μm, while it shows zero reflectance for $\lambda = 15$ μm. There are N = 5 design parameters ($\Lambda$, F, $d_g$, $d_L$, $d_M$), and the search space is chosen as: $\Lambda$(μm) ∈ {5.0,8.0}, F ∈ {0,1.0}, $d_g$(μm) ∈ {0,4.0}, $d_L$(μm) ∈ {0,4.0}, $d_M$(μm) ∈ {0,4.0}. The fitness function is defined as:

$$FF = \left(\frac{1}{N_\lambda} \sum_{i=1}^{N_\lambda} \left(R_{TE,desired}(\lambda_i) - R_{TE,designed}(\lambda_i)\right)^2\right)^{\frac{1}{2}}, \quad (1)$$

where $N_\lambda$ is the number of wavelength points, here set to $N_\lambda = 5$. Upon 260 iterations, we obtain an optimal filter with FF = 0.0445. The optimized design parameters are: $\Lambda$ = 7.84 μm, F = 0.53, $d_g$ = 0.54 μm, $d_L$ = 2.92 μm and $d_M$ = 4.0 μm. Figure 2(a) presents the spectral response of the emitter for TE polarization. As specified, we achieve a very narrow emissivity/absorption peak at 15 μm. As a note, a computer with an Intel Core i5-11500 @ 2.70GHz processor and 16GB of RAM has been employed for this calculation.

As mentioned earlier, GMRs in an all-dielectric structures exhibit sharp peaks in the reflection spectrum. This is a result of destructive interference in the transmission medium that gives rise to perfect reflection. By incorporating an anisotropic α-MoO$_3$ layer as a polarization-dependent reflector in the dielectric GMR structure, the boundary conditions change, thus the transmission is zero within the RB along the *y* direction. As a result, an absorption-like feature appears in the reflection spectrum, which is the origin of the narrowband thermal emission that we demonstrate. Figure 2(b) shows the electric field distribution (real($E_y$)) at λ=15 μm, where the GMR field profile in the ZnSe layer/waveguide is clearly visible. This is similar to the behavior of metal-assisted GMR structures [32].

Next, we provide a deeper insight into the device spectral behavior. First, the absorption/emissivity peak in Fig. 2(a) depends on the angle of incidence. In Fig. 3(a), we show reflectance as a function of wavelength and angle of incidence for TE polarization, which gives information about the photonic band structure. At $\theta = 0$, only one resonance is present at $\lambda = 15\mu m$ as expected, while by deviating from normal incidence, two resonances appear due to induced asymmetry [33]. By examining the electric field distributions, all resonances at normal and oblique incidences exhibit $TE_0$ GMR modal features. At $\theta = 0$, band closure occurs, which is also known as Dirac point [34]. This is the point at which the structure is almost symmetric and F~0.5. Figure 3(b) shows the effect of the $\alpha$-MoO$_3$ layer thickness, all other parameters being kept to their optimized values. As a reference, when $d_{MoO3} = 0$, there is a standard GMR reflection peak in the spectrum as expected. As the thickness increases, absorption features emerge and shift towards shorter wavelengths as $d_{MoO3}$ increases. For thicknesses beyond 1 µm, no further changes in reflectance is observed.

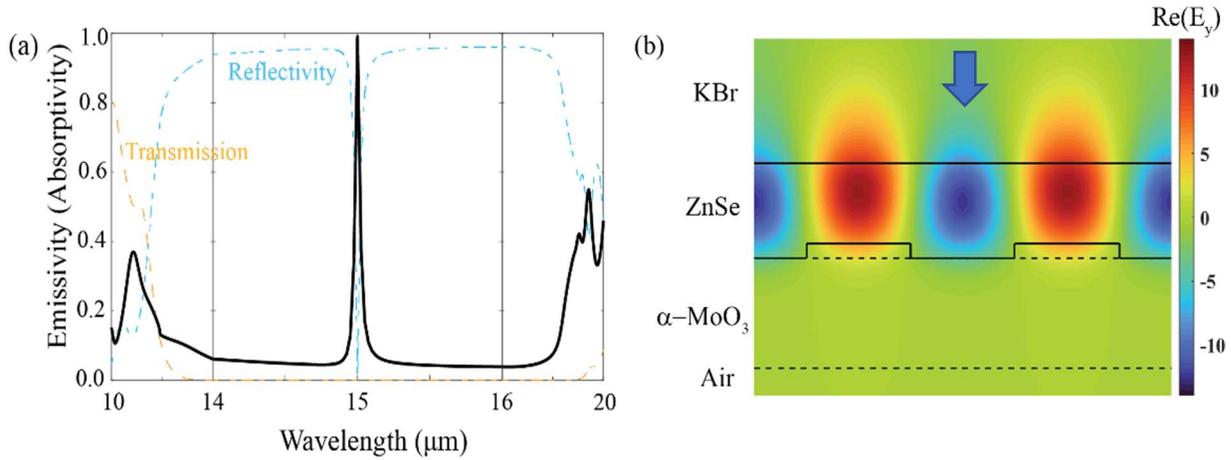

Fig. 2. (a) The spectral response of the designed emitter for TE polarization ($\Lambda = 7.84$ µm, F = 0.53, $d_g = 0.54$ µm, $d_L = 2.92$ µm and $d_M = 4.0$ µm), (b) Electric field distribution at $\lambda$=15 µm (incidence is from the top (substrate side)).

Figure 4(a) shows the color-coded reflectance, R($\lambda$,$d_L$) map for TE polarization. As shown, it represents the modal behavior of leaky modes: at $d_L = 2.92$ µm, only $TE_0$ waveguide mode is present (see also Fig. 2(b)), while by increasing the layer/waveguide thickness, for example at $d_L = 7.19$ µm, another resonance corresponding to the $TE_1$ mode appears. The reflectance response of the element at this point is shown in Fig. 4(b). In addition, the electric field distribution at $\lambda = 14.41$ µm is shown as inset of Fig. 4(b), which shows higher GMR mode of $TE_1$. Figures 3(b) and 4 demonstrate how the GMR modes are excited in ZnSe layer under TE illumination: For TE polarization no phonon polariton mode is excited and observable in the $\alpha$-MoO$_3$ layer.

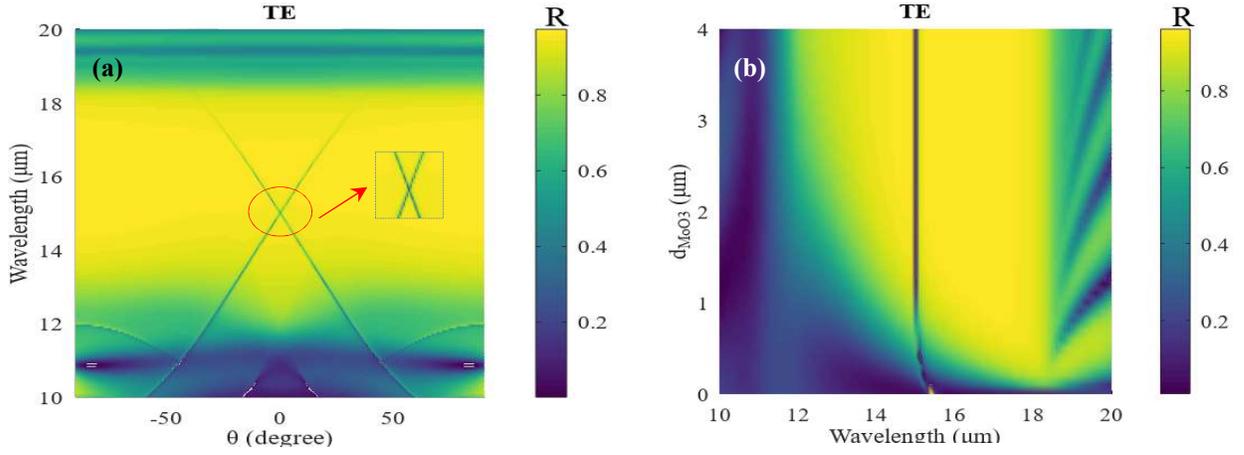

Fig. 3. (a) The color-coded R(θ,λ) map for TE polarization. (b) The effect of α-MoO$_3$ layer thickness on spectral response of the element under normal TE illumination.

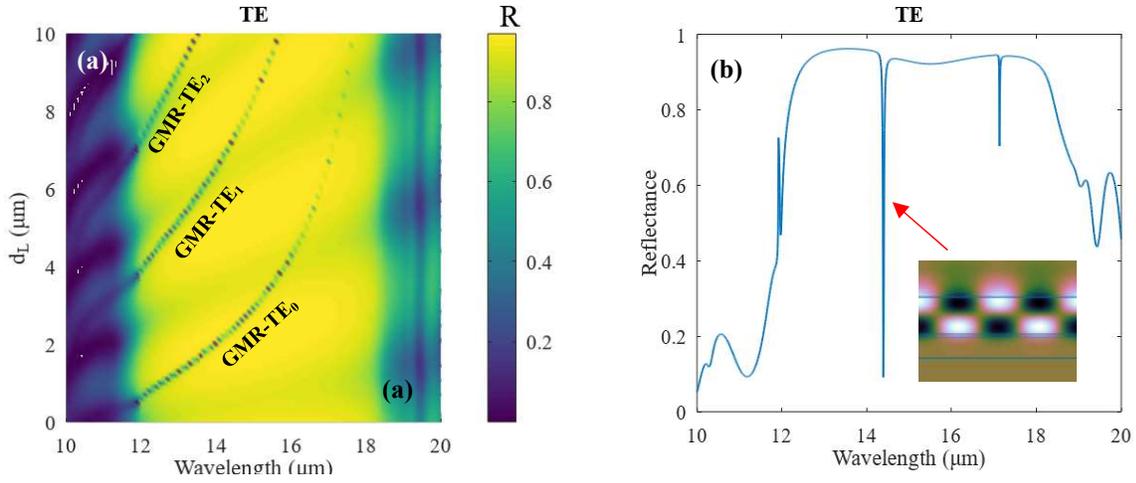

Fig. 4. (a) The color-coded R(λ,d$_L$) map for TE polarization. (b) The reflectance response of the element at d$_L$ = 7.19 μm. The electric field distribution at λ = 14.41 μm is shown as inset.

Narrowband emitters can also be designed for TM polarized illumination, and exhibit narrow absorption-like spectral features in the RB$_x$ band. The second device has been designed by PSO to exhibit a resonance at 11.5 μm wavelength (almost at the center of the RB$_x$). The search space is the same as the one for TE polarization, but the fitness function and target reflection spectrum are modified. After 232 iterations, we reach an optimal design with FF = 0.0788. The optimized parameters are: Λ = 5.74 μm, F = 0.71, d$_g$ = 3.66 μm, d$_L$ = 3.97 μm and d$_M$ = 4.0 μm. Figure 5(a) presents the spectral response of the device for normal incidence and TM polarization. We observe a sharp emissivity peak at 11.5 µm, as targeted. The magnetic field distribution profile (real($H_y$)) at $\lambda = 11.5$ µm is shown in Fig. 5(b). This demonstrates a standing wave within the ZnSe layer, corresponding to the GMR mode of $TM_0$. The field features in the grating and α-MoO3 layer are a hybridization between α-MoO3 hyperbolic modes (phonon polaritons) and guided modes of that layer [35, 36].

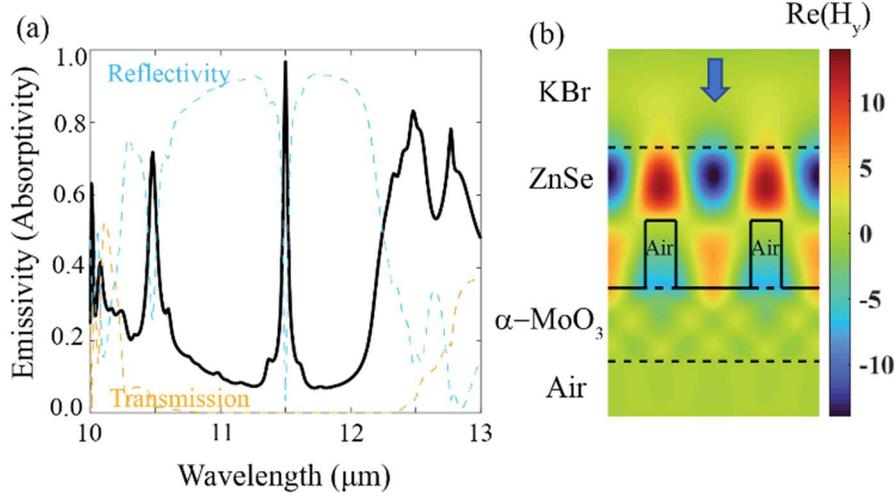

Fig. 5. (a) The spectral response of the device under normal incidence and TM polarization ($\Lambda = 5.74$ µm, F = 0.71, $d_g = 3.66$ µm, $d_L = 3.97$ µm and $d_M = 4.0$ µm). (b) The magnetic field distribution profile at $\lambda = 11.5$ µm (incidence is from the top (substrate side)).

To further investigate this designed element, the behavior of the device under oblique incidence and its band structure has been studied by presenting the $R(\theta, \lambda)$ map for TM polarization in Fig. 6(a). As seen, in contrast to TE case, there are complex spectral features in the band structure of this designed element. This figure shows a complex band structure which is a result of hybridization of the GMR and phonon polariton modes [35-37]. Under normal incidence, the resonance, also shown in Fig. 5(a), is visible and both GMR features in the ZnSe layer/waveguide and phononic features in the α-MoO3 layer are observable. For further insight, Fig. 6(b) shows the band structure of the device when $d_{MoO3} = 0$, which shows a standard band structure of a symmetric GMR device. Also, Fig. 6(c) illustrates the band structure of the device when frequency dispersion is absent, in other words when we consider that the diagonal elements of α-MoO3 are frequency-independent and real values equal to $\varepsilon_{i,\infty}$ ($i = x,y,z$); The anisotropic dielectric and GMR band features are clearly seen. By adding the actual α-MoO3 layer (including frequency dispersion [28]) as shown in Fig. 6(a), phononic waveguide mode is introduced, and a high reflection background from the Restrahlen band ($RB_x$) arises. The result is the emergence of the narrow absorption feature at λ = 11.5 µm at vicinity of normal incidence, which enables both directional and narrowband emission simultaneously.

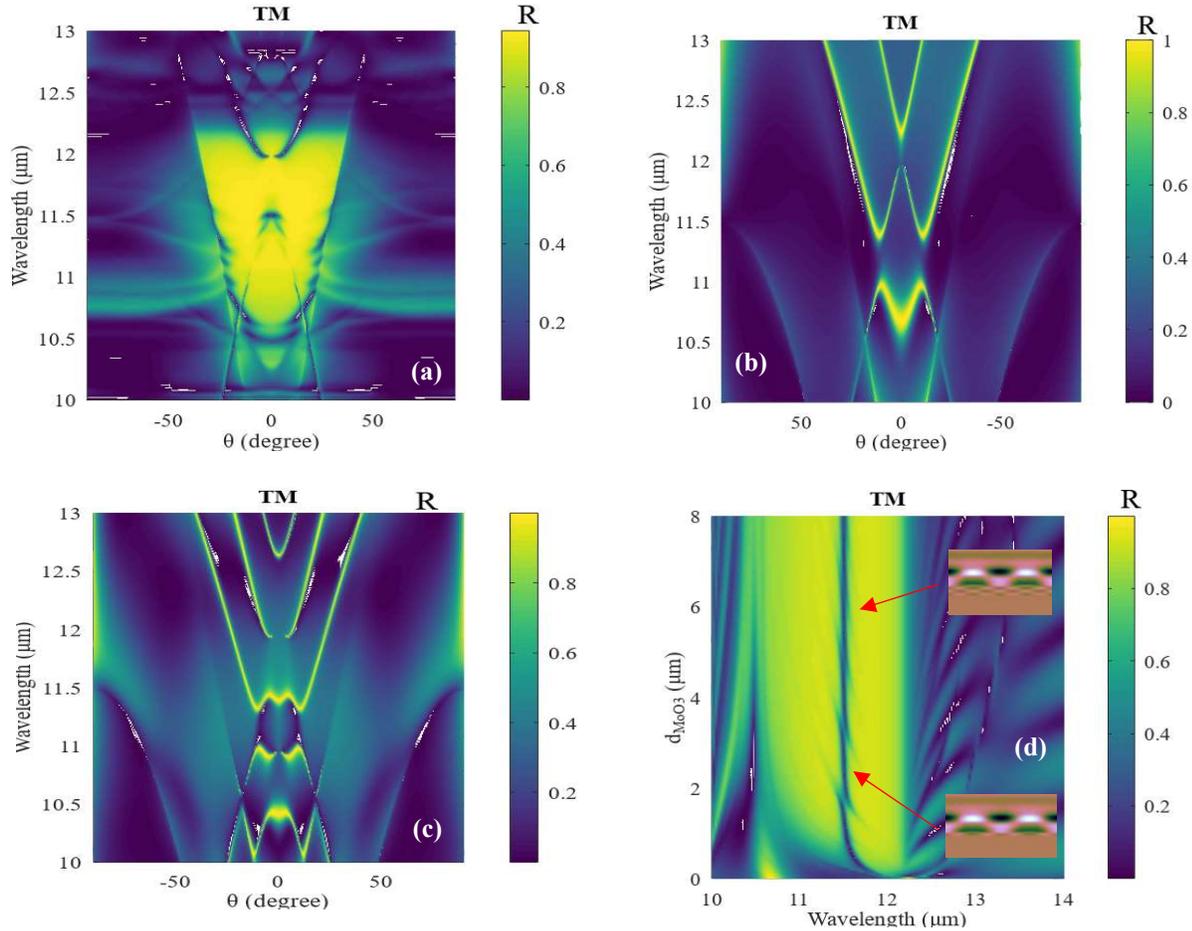

Fig. 6. (a) The color-coded R(θ, λ) map for TM polarization. (b) The band structure of the device when $d_{MoO3}$ = 0, (c) The band structure of the device when the diagonal elements of α-MoO₃ are frequency-independent and are real values equal to $\varepsilon_{i,\infty}$ (i = x,y,z), (d) The effect of α-MoO₃ layer thickness on spectral response of the element under normal TM illumination.

Figure 6(d) shows the effect of α-MoO₃ layer thickness. As seen, when $d_{MoO3}$ = 0, there is a standard GMR reflection peak in the spectrum (at 12.23 µm) and while the thickness increases, the absorption feature prevails. Although the position of resonance does not change for fairly thick α-MoO₃ layer, the crossings arise due to interacting phononic modes. In contrast to the TE polarization case, both Figs. 6(a) and 6(d) show complex spectral features, which is related to excitation and hybridization of both GMR and phonon polaritons for different angles of incidence and $d_{MoO3}$.

Figure 7(a) shows the color-coded reflectance $R(\lambda,d_L)$ map for TM polarization which is representative for the behavior of the GMR modes. As previously seen in Fig. 5(b), at $d_L$ = 3.97 µm, a GMR mode is present in ZnSe layer, while by increasing ZnSe layer/waveguide thickness, for example to $d_L$ = 8.8 µm, another resonance corresponding to the higher GMR mode appears. The reflectance response of the element at this point is shown in Fig. 7(b) and as seen the GMR mode has $TM_1$ features. In addition, the magnetic field distribution at $\lambda$ = 11.54 µm is shown as inset in Fig. 7(b).

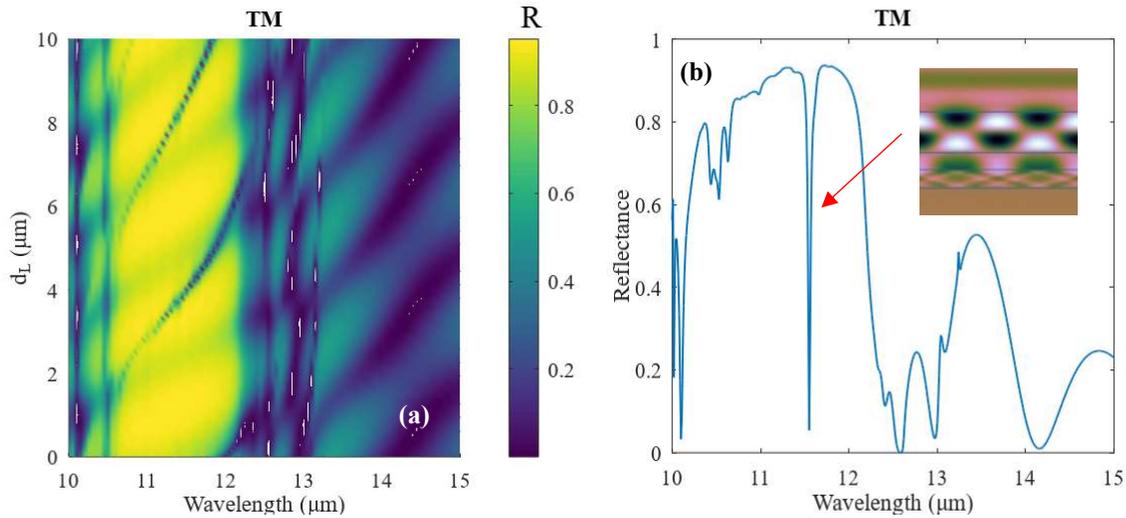

Fig. 7. (a) The color-coded $R(\lambda,d_L)$ map for TM polarization, (b) The reflectance response of the element at $d_L$ = 8.8 µm. The magnetic field distribution at $\lambda$ = 11.54 µm is shown as inset.

By rotating the elements (or the α-MoO$_3$ flake) by 90° (or changing the azimuthal angle from 0 to 90°), the spectral response switch between TE and TM polarization responses in both aforementioned designed 1D elements as expected. This operation needs mechanical rotation of the element/flake. In addition, in each of the designed elements, the target spectrum was considered for a specific polarization. Therefore, although spectral switching takes place by rotation, only one spectrum is desired since the device has been optimized for one polarization, and the other spectrum is just the unoptimized response of the designed element to the other polarization.

**Design of 2D narrowband emitters**

To realize a device with the desired dichroic spectral response, we consider a two-dimensional structure and optimize it by PSO to simultaneously exhibit desired spectral responses for TE and TM polarizations, without rotation. Figure 8(a) illustrates the schematic of the proposed element. In this case, there are N = 8 design parameters ($\Lambda_x$, $\Lambda_y$, $F_x$, $F_y$, $d_g$, $d_L$, $d_M$) for PSO optimization. The desired spectra and fitness function are chosen such that, under normal incidence, the device response for TE and TM polarization corresponds to the ones belonging to the first and second previously designed elements (For TE, $\lambda_e$ = 15.0 μm and for TM, $\lambda_e$ = 11.50 μm), respectively. Upon 3048 iterations, we obtain an optimal filter with FF = 0.1590. The optimized design parameters are: $\Lambda_x$ = 7.57 μm, $\Lambda_y$ = 5.50 μm, $F_x$ = 0.58, $F_y$ = 0.76, $d_g$ = 0.83 μm, $d_L$ = 3.27 μm and $d_M$ = 2.22 μm. Figures 8(b) and 8(c) show the spectral response of the designed 2D element for TE and TM polarizations, respectively. As seen and expected from PSO design, there are narrowband emissions at 15 μm and 11.5 μm for TE and TM polarizations, respectively. Figures 8(d) and 8(e) illustrate the R($\lambda$, $\theta$) maps (as a part of element's band structures) for TE and TM polarizations. As seen, there are similarities between features in these figures and Figs. 3(a) and 6(a), respectively, except that in the 2D structure, these band structures are simultaneously present and can be switched just by changing the polarization.

This element can be employed as a polarization-dependent dichroic absorption filter or emitter. The response of the device can be considered as a combination of the TE and TM polarization responses similar to the procedure reported in [38] except that anisotropy of the α-MoO₃ should be taken into account. This element can be a potential candidate for applications such as thermal sources for applications that need to detect two distinct signatures at the same time, thermal sources with tunable emission wavelengths for different polarizations, and polarization-dependent camouflage.

**Conclusions**

In summary, by combining an anisotropic layer of α-MoO₃ with a GMR configuration, we designed 1D and 2D narrowband, polarization-dependent infrared emitters by particle swarm optimization (PSO) within the wavelength range of 10-20 μm. We show that well-defined spectral absorption/emissivity features are present for both TE and TM polarizations.

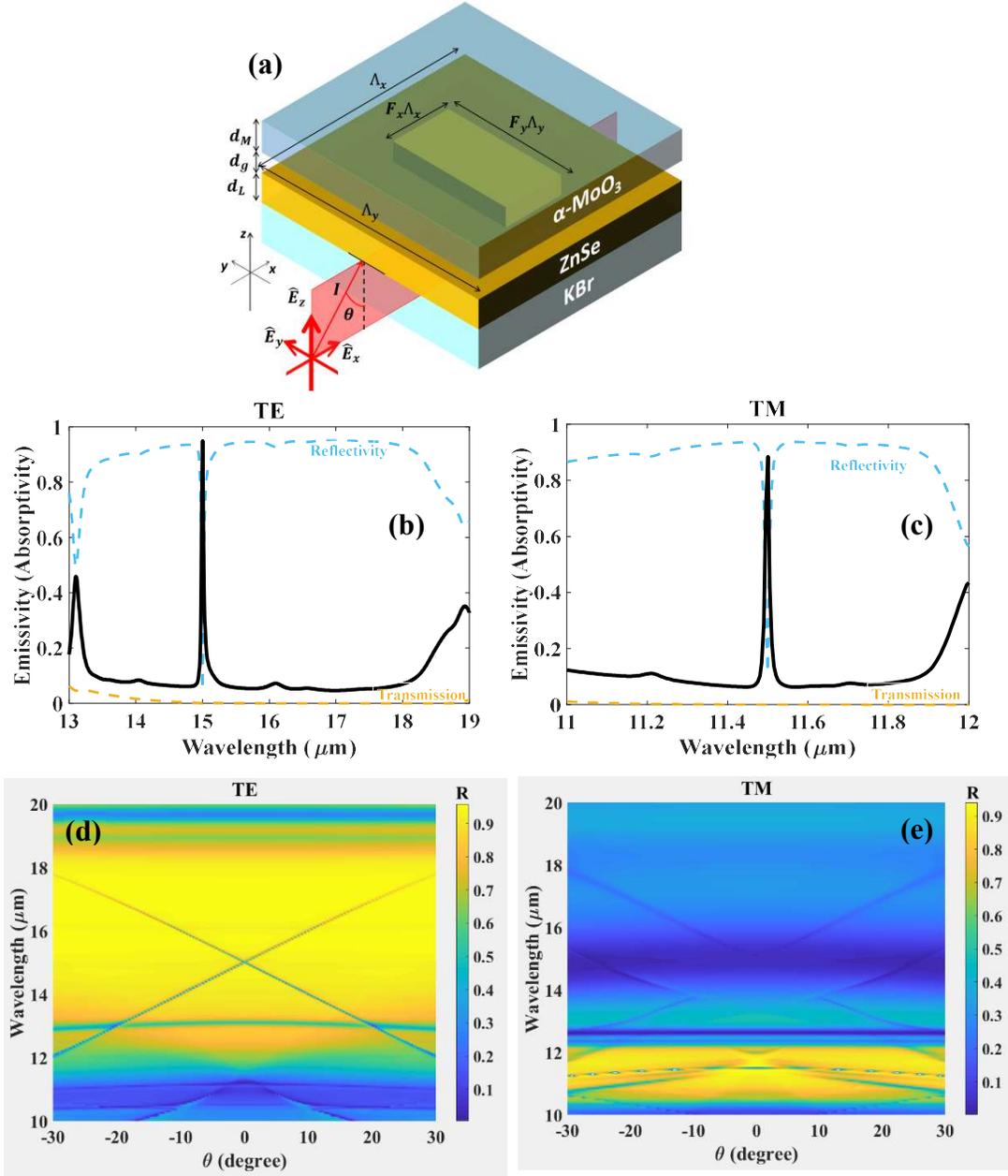

Fig. 8. (a) Schematic view of the proposed 2D element. The periods of grating in x and y directions, grating filling factors in x and y directions, thicknesses of grating, ZnSe layer and α-MoO₃ layer are shown by $\Lambda_x$, $\Lambda_y$, $F_x$, $F_y$, $d_g$, $d_L$, and $d_M$, respectively. Illumination is from the substrate side. (b) Spectral response of the designed 2D element for TE polarization. (c) Spectral response of the designed 2D element for TM polarization. (d) R(λ, θ) map for TE polarization. (e) R(λ, θ) map for TM polarization.

The band structure of the designed emitters under TM polarization exhibits complex behavior unlike that for TE polarization. This arises from the guided-mode resonance and phonon polariton modes hybridization. We have also studied the physical behavior of the designed elements and investigated the tunability of the narrowband emission by angle of incidence and structural parameters. The designed two-dimensional emitter achieves narrowband emission at distinct frequencies for TE and TM polarizations. These results are relevant for applications in active and passive photonic elements in mid- and longwave IR bands.

**Acknowledgements:** M. Shokooh-Saremi would like to acknowledge the support from ICFO through Severo Ochoa Senior Visiting Scientists program. M. Giteau acknowledges financial support from the Severo Ochoa Excellence Fellowship. We acknowledge financial support from the Department of Research and Universities of the Generalitat de Catalunya (2021 SGR 01443). G.T.P. acknowledges funding from "la Caixa" Foundation (ID 100010434), from the PID2021-125441OA-I00 project funded by MCIN/AEI/10.13039/501100011033/FEDER, UE, and from the European Union's Horizon 2020 research and innovation programme under the Marie Skłodowska-Curie Grant Agreement No. 847648. The fellowship code is LCF/BQ/PI21/11830019. This work is part of the R&D project CEX2019-000910-S, funded by MCIN/AEI/10.13039/501100011033/, from Fundació Mir-Puig, and from Generalitat de Catalunya through the CERCA program.

**Data availability:** Data underlying the results presented in this paper are not publicly available at this time but may be obtained from the authors upon reasonable request.

**Disclosures:** The authors declare no conflicts of interest.

**References**

[1] G. T. Papadakis, M. Orenstein, E. Yablonovitch, and S. Fan, "Thermodynamics of light management in near-field thermophotovoltaics," Physical Review Applied, vol. **16**, 064063, 2021.

[2] Q. Wang, G. Hou, Y. Zhu, T. Sun, J. Xu, and K. Chen, "Nanolayered wavelength-selective narrowband thermal emitters for solar thermophotovoltaics," ACS Applied Nano Materials, vol. 5, 13455−13462, 2022.

[3] S. Fan and W. Li, "Photonics and thermodynamics concepts in radiative cooling," Nature Photonics, vol. 16, no. 3, 182–190, 2022.

[4] A. Lochbaum, A. Dorodnyy, U. Koch, S. M. Koepfli, S. Volk, Y. Fedoryshyn, V. Wood, and J. Leuthold, "Compact mid-infrared gas sensing enabled by an all-metamaterial design," Nano Letters, vol. 20, 4169-4176, 2020.

[5] H.-S. Lee, J.-S. Cha, J.-Y. Jin, Y.-J. Ko, and T.-Y. Seong, "Fabrication of thermally stable mid-infrared optical filters using tantalum microdisc array," Applied Physics Letters, vol. 121, 161101, 2022.

[6] M. Dai, C. Wang, B. Qiang, F. Wang, M. Ye, S. Han, Y. Luo, and Q. J. Wang, "On-chip mid-infrared photothermoelectric detectors for full-Stokes detection," Nature Communications, vol. 13, 4560, 2022.

[7] A. John-Herpin, A. Tittl, L. Kühner, F. Richter, S. H. Huang, G. Shvets, S.-H. Oh, and H. Altug, "Metasurface-enhanced infrared spectroscopy: An abundance of materials and functionalities," Advanced Materials, vol. 35, 2110163, 2022.

[8] M. Muhiyudin, D. Hutson, D. Gibson, E. Waddell, S. Song, and S. Ahmadzadeh, "Miniaturised infrared spectrophotometer for low power consumption multi-gas sensing," Sensors, vol. 20, 3843, 2020.

[9] A. Lochbaum, Y. Fedoryshyn, A. Dorodnyy, U. Koch, C. Hafner, and J. Leuthold, "On-chip narrowband thermal emitter for mid-IR optical gas sensing,". ACS Photonics, vol. 4, no. 6, 1371–1380, 2017.

[10] D. G. Baranov, Y. Xiao, I. A. Nechepurenko, A. Krasnok, A. Alù, and M. A. Kats, "Nanophotonic engineering of far-field thermal emitters," Nature Materials, vol. 18, no. 9, 920–930, 2019.

[11] A. Hessel and A. Oliner, "A new theory of Wood's anomalies on optical gratings," Applied Optics, vol. 4, 1275–1297, 1965.

[12] S. S. Wang and R. Magnusson, "Theory and applications of guided-mode resonance filters," Applied Optics, vol. 32, 2606–2613, 1993.

[13] G. Quaranta, G. Basset, O. J. F. Martin, and B. Gallinet, "Recent advances in resonant waveguide gratings," Lasers & Photonics Reviews, vol. 12, 1800017, 2018.

[14] M. He, T. G. Folland, J. Duan, P. Alonso-González, S. De Liberato, A. Paarmann, and J. D. Caldwell, "Anisotropy and modal hybridization in infrared nanophotonics using low-symmetry materials," ACS Photonics, vol. 9, 1078-1095, 2022.

[15]  S. Moon, J. Kim, J. Park, S. Im, J. Kim, I. Hwang, and J. K. Kim, "Hexagonal boron nitride for next-generation photonics and electronics," Advanced Materials, vol. 35., 2204161, 2023.

[16]  D. Andres-Penares, M. Brotons-Gisbert, C. Bonato, J. F. Sánchez-Royo, and B. D. Gerardot, "Optical and dielectric properties of $MoO_3$ nanosheets for van der Waals heterostructures," Applied Physics Letters, vol. 119, 223104, 2021.

[17]  D. N. Basov, M. M. Fogler, and F. J. García de Abajo, "Polaritons in van der Waals materials," Science, vol. 354, no. 6309, aag1992, 2016.

[18]  W. Ma, P. Alonso-González, S. Li, A. Y. Nikitin, J. Yuan, J. Martín-Sánchez, J. Taboada-Gutiérrez, I. Amenabar, P. Li, S. Vélez, C. Tollan, Z. Dai, Y. Zhang, S. Sriram, K. Kalantar-Zadeh, S.-T. Lee, R. Hillenbrand, and Q. Bao, " In-plane anisotropic and ultra-low-loss polaritons in a natural van der Waals crystal," Nature, vol. 562, 557-562, 2018.

[19]  Z. Zheng, J. Chen, Y. Wang, X. Wang, X. Chen, P. Liu, J. Xu, W. Xie, H. Chen, S. Deng, and N. Xu, "Highly confined and tunable hyperbolic phonon polaritons in van der Waals semiconducting transition metal oxides," Advanced Materials, vol. 30, 1705318, 2018.

[20]  J. D. Caldwell, L. Lindsay, V. Giannini, I. Vurgaftman, T. L. Reinecke, S. A. Maier, and O. J. Glembocki, "Low-loss, infrared and terahertz nanophotonics using surface phonon polaritons," Nanophotonics, vol. 4, 44-68, 2015.

[21]  G. Hu, J. Shen, C.-W. Qiu, A. Alù, and S. Dai, "Phonon polaritons and hyperbolic response in van der Waals materials," Advanced Optical Materials, vol. 8, 1901393, 2020.

[22]  M. Ye, B. Qiang, S. Zhu, M. Dai, F. Wang, Y. Luo, Q. Wang, and Q. J. Wang, "Nano-optical engineering of anisotropic phonon resonances in a hyperbolic α-MoO3 metamaterial," Advanced Optical Materials, vol. 10, 2102096, 2022.

[23]  G. Zheng, X. Zou, F. Xian, and L. Xu, "Cavity-mediated interactions between the guided mode resonance and surface phonon polaritons in a hybrid photonic–phononic system," Applied Physics Express, vol. 11, 082001, 2018.

[24]  S. Pechprasarn, S. Learkthanakhachon, G. Zheng, H. Shen, D. Y. Lei, and M. G. Somekh, "Grating-coupled Otto configuration for hybridized surface phonon polariton excitation for local refractive index sensitivity enhancement," Optics Express, vol. 24, 19517-19530, 2016.

[25]  K. Ito, T. Matsui, and H. Iizuka, "Thermal emission control by evanescent wave coupling between guided mode of resonant grating and surface phonon polariton on silicon carbide plate," Applied  Physics Letters, vol. 104, 051127, 2014.

[26] J. Kennedy and R. Eberhart, "Particle swarm optimization," Proceedings of ICNN'95 - International Conference on Neural Networks, Perth, WA, Australia, vol. 4, pp. 1942-1948, 1995.

[27] M. Shokooh-Saremi and R. Magnusson, "Particle swarm optimization and its application to the design of diffraction grating filters," Optics Letters, vol. 32, pp. 894-896, 2007.

[28] G. Álvarez-Pérez, T. G. Folland, I. Errea, J. Taboada-Gutiérrez, J. Duan, J. Martín-Sánchez, A. I. F. Tresguerres-Mata, J. R. Matson, A. Bylinkin, M. He, W. Ma, Q. Bao, J. I. Martín, J. D. Caldwell, A. Y. Nikitin, and P. Alonso-González, "Infrared permittivity of the biaxial van der Waals semiconductor α-MoO3 from near- and far-field correlative studies," Advanced Materials, vol. 32, 1908176, 2020.

[29] M. G. Moharam, E. B. Grann, D. A. Pommet and T. K. Gaylord, "Formulation for stable and efficient implementation of the rigorous coupled-wave analysis of binary gratings", Journal of the Optical Society of America A, vol. 12, pp. 1068-1076, 1995.

[30] L. Li, "New formulation of the Fourier modal method for crossed surface-relief gratings," Journal of the Optical Society of America A, vol. 14, 2758-2767, 1997.

[31] J. P. Hugonin and P. Lalanne, RETICOLO software for grating analysis, Institut d'Optique, Orsay, France, 2005, arXiv:2101:00901.

[32] S.-F. Lin, C.-M. Wang, T.-J. Ding, Y.-L. Tsai, T.-H. Yang, W.-Y. Chen, and J.-Y. Chang, "Sensitive metal layer assisted guided mode resonance biosensor with a spectrum inversed response and strong asymmetric resonance field distribution," Optics Express, vol. 20, pp. 14584-14595, 2012.

[33] Y. Ding and R. Magnusson, "Use of nondegenerate resonant leaky modes to fashion diverse optical spectra," Optics Express, vol. 12, pp. 1885-1891, 2004.

[34] T. Jacqmin, I. Carusotto, I. Sagnes, M. Abbarchi, D. D. Solnyshkov, G. Malpuech, E. Galopin, A. Lemaître, J. Bloch, and A. Amo, "Direct observation of Dirac cones and a flatband in a honeycomb lattice for polaritons," Physical Review Letters, vol. 112, 116402, 2014.

[35] S. Foteinopoulou, G. C. R. Devarapu, G. S. Subramania, S. Krishna, and D. Wasserman, "Phonon-polaritonics: Enabling powerful capabilities for infrared photonics," Nanophotonics, vol. 8, no. 12, pp. 2129-2175, 2019.

[36] N. C. Paßler, "Phonon polaritons in polar dielectric heterostructures," Ph.D. Dissertation, Freie Universität Berlin, 2020.

[37] K. Y. Lee, K. W. Yoo, Y. Choi, G. Kim, S. Cheon, J. W. Yoon, and S. H. Song, "Topological guided-mode resonances at non-Hermitian nanophotonic interfaces," Nanophotonics, vol. 10, pp. 1853–1860, 2021.

[38] Y. H. Ko, M. Shokooh-Saremi, and R. Magnusson, "Modal processes in two-dimensional resonant reflectors and their correlation with spectra of one-dimensional equivalents," IEEE Photonics Journal, vol. 7, 4900210, 2015.